\documentclass[twocolumn,showpacs,preprintnumbers]{revtex4}

\usepackage{graphicx}
\usepackage{dcolumn}
\usepackage{bm}
\usepackage{color}
\usepackage{amsmath}
\usepackage{amsfonts}
\usepackage[pdftex]{epsfig}
\usepackage{epstopdf}
\usepackage{mhchem}
\begin{document}

\title{Reentrant superconductivity in a composite formed by  YBa$_2$Cu$_3$O$ _{7-\delta}$ and Ammonium Terbium Oxalate.}

\author{Rodolfo E. L\'{o}pez-Romero and Dulce Y. Medina}
\affiliation{Divisi\'{o}n de Ciencias B\'{a}sicas e Ingenier\'{i}a, Universidad Aut\'{o}noma Metropolitana-Azcapotzalco, Av. San Pablo No 180, Col. Reynosa-Tamaulipas, C.P. 02200 M\'{e}xico D.F., M\'{e}xico.}

\author{R. Escudero}
\affiliation{Instituto de Investigaciones en Materiales, Universidad Nacional Aut\'{o}noma de M\'{e}xico. A. Postal 70-360. M\'{e}xico, D.F.}

\date{\today}

\begin{abstract}
We present a study of  reentrant behaviour in a composite formed by a Hight-T$ _{c} $  superconductor, YBa$_2$Cu$_3$O$ _{7-\delta}$ and Ammonium Terbium Oxalate, Tb(H$_2$O)(C$ _2$O$_4$)$_2$ $\cdot $NH$_4$. The composite has a transition temperature about 92 K, and it presents a reentrant behaviour resulting of the coexistence between superconductivity and magnetism. According to this study the values and shape of the critical magnetic fields were dramatically reduced in a similar form as in other known reentrant superconductors.  

\end{abstract}


\maketitle

\section{Introduction}
Among the many  interesting characteristics of some  superconductors, reentrant behaviour in superconductivity is a quite important topic of study. Some   superconductors  displaying this behaviour are those that   have a magnetic ion  in its crystal structure. A classical  example very  well known is ErRh$ _{4} $B$ _{4} $, it is a reentrant compound   found by  Ferting many years ago \cite{fer}.  Many others were  discovered with this characteristic  \cite{ishi, eis, pena}. 
The physics related to the reentrant  process occurs because  the competition between  superconductivity and magnetism.  This phenomenon   is  actual and  important topic  of  study,  it was first  addressed by Ginzburg \cite{ginz}, he studied  the possibility of  coexistence of   superconductivity and  ferromagnetic materials.  Quite important to mention is that in BCS theory these two phenomena are antagonist and mutually excludents. In   BCS   the phenomenon  arises by  the creation of   Cooper pairs   mediated by  virtual phonons connecting the  two electrons and resulting in  ordered   antiparalel spins;  the BCS manner of  Cooper pairs formation. If paramagnetic impurities are in the  superconducting  bulk those  magnetic ions  could  break Cooper pairs, destroying superconductivity, depending on the strength of the magnetic interactions, then a new  spins ordering will be arises. This  new arrange causes  breaking of pairs because a type of ferromagnetic order;  parallel spins,  and the resulting  parallel configuration will  bring the destruction of the superconducting state; thus, reentrant superconducting behaviour will arise.

In a possible reentrant superconductor,   if the intensity of the  magnetic interactions are  strong enough  will affect and influence   the behavior of the superconducting state. Accordingly,  the magnetic  interactions could give additional information to  have a  better understanding of a  superconductor and in that manner to obtain more insight into  the microscopic mechanisms and electronic properties. Then  some of the basic  phenomena  related to the coupling and formation of the Cooper pairs  will be related to this  elementary excitations. 

The reentrant behavior in superconductors   is well know since early times. As mentioned it may occurr when a superconductor is doped with paramagnetic impurities, the interplay results in pair breaking and  total destruction of superconductivity. This phenomenon  has been explained by Abrikosov and Gor'kov \cite{abr}, they study the interplay introducing the form that magnetic ions play an important role in the exchange interaction of both conducting electrons an magnetic ions \cite{maple, gupta, eis}.

As was  mentioned,  in BCS theory the two phenomenons  are mutually excludents.  In many  compound, however  recently it has been observed that  more  types of  superconductors involved different forms of  pair formation,  meaning    that the nature of pair coupling will  be  different to the mediated by phonons. Thus, the  superconducting state  will be different as in the BCS model.  Among those new materials, the Fe-Based compounds are   a type of the new materials \cite{liu}.  New studies by  different researchers in the superconducting field  have discovered many new  example of these materials not following BCS theory,  with different manners to form the Cooper pairs mediated with different processes of coupling.

The effect of the influences of a magnetic field in a superconductor is most notable  at  temperatures below the superconducting  transition. In this situation the influence  of the magnetism is quite clear because  the transition temperature decreases due to the diminution of  the number of  Cooper  pairs. The most clear consequence of these  interactions  between  superconductivity and magnetic field   results in the  reentrant superconducting behavior. These interactions create a competition between both, magnetism and superconductivity, growing and reinforcing with decreasing temperature, although  those changes will be  at different  rates. Depending of the strengths  and magnitude of those  processes,  one or the other could be predominant. In some cases   the superconducting state could be destroyed or simply  reduced. In other situations  the two interacting energies,  one  to create Cooper pairs,  the other to energize magnetic  interchanges could be in equilibrium \cite{maple}.  Those processes are very dependent on temperature. Normally, the energy for  Cooper pairs formation  increases when  the temperature decreases, but this is not always the case for the ferromagnetic  interactions, it can be decreases,  or increases depending on many factors; i.e.  type of crystal structure, type of different transitions, and or  other modification related with temperature. Accordingly in some materials the superconducting pair  will be only reduced. 
The reentrant characteristic in  superconducting materials  was for the  first time  observed by Ferting \cite{fer} in the compound $ErRh_4B_4$. Posteriorly it was seen  in other compounds as $HoMo_6Se_8$ \cite{ishi}, also  in $HoNi_2B_2C$ \cite{eis, pena}, and in others. 

In this research we present a study of the reentrant behaviour in a composite formed by a ceramic superconductor and a luminescent material.  The experiments were carried out using YBa$_2$Cu$_3$O$_{7-\delta}$ particles and two luminescent materials; Tb(H$_{2}$O)(C$_{2}$O$_{4}$)$_{2}$ $\cdot$ NH$_{4}$, or Y$_{2}$O$_{3}$:Eu$^{3+}$. Two composites to different proportions in weight between its components were prepared. Both  were characterized by magnetic susceptibility  as  function of temperature and magnetic field applied. Only the composite YBCO - Tb(H$_{2}$O)(C$_{2}$O$_{4}$)$_{2}$ $\cdot$ NH$_{4}$ presents reentrant superconductivity. The critical fields were evaluated in the composite 10\% superconductor - 90\% Tb oxalate and we found a drastic reduction of them due to the coexistence between superconductivity and magnetism. Initially idea of the research was to  study the  influence of the luminescent materials into  the superconductor under UV illumination at low temperatures.

\begin{figure}[h]
	\centering
		\includegraphics[width=0.40\textwidth]{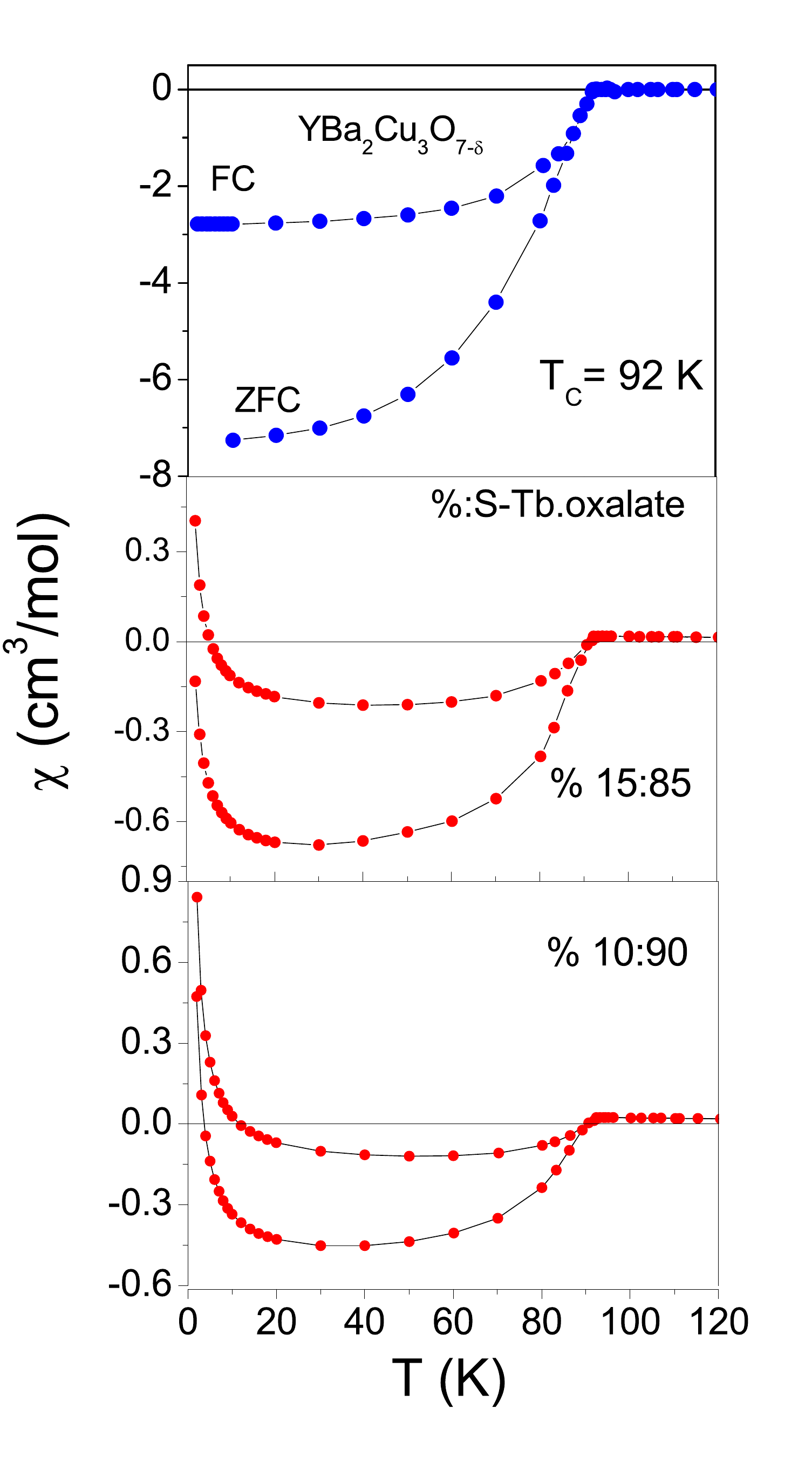}
		   \caption{(Color on-line) Three graphics of susceptibility,  $\chi(T)$. Top figure is the characteristic  of the pure  superconductor,  with $T_C = 92$ K   determined by measurements  in ZFC and FC modes and  $H = 100$ Oe. The other two  figures show the reentrant behaviour in the composite  superconductor - Tb oxalate at two proportions of superconducting material; 10 and 15 \%. Note the reentrance  behaviour as the \% of superconductor and Tb oxalate decreases and increases respectively. }
	\label{Fig1.pdf}
\end{figure}

\begin{figure}[h]
	\centering
		\includegraphics[width=0.5\textwidth]{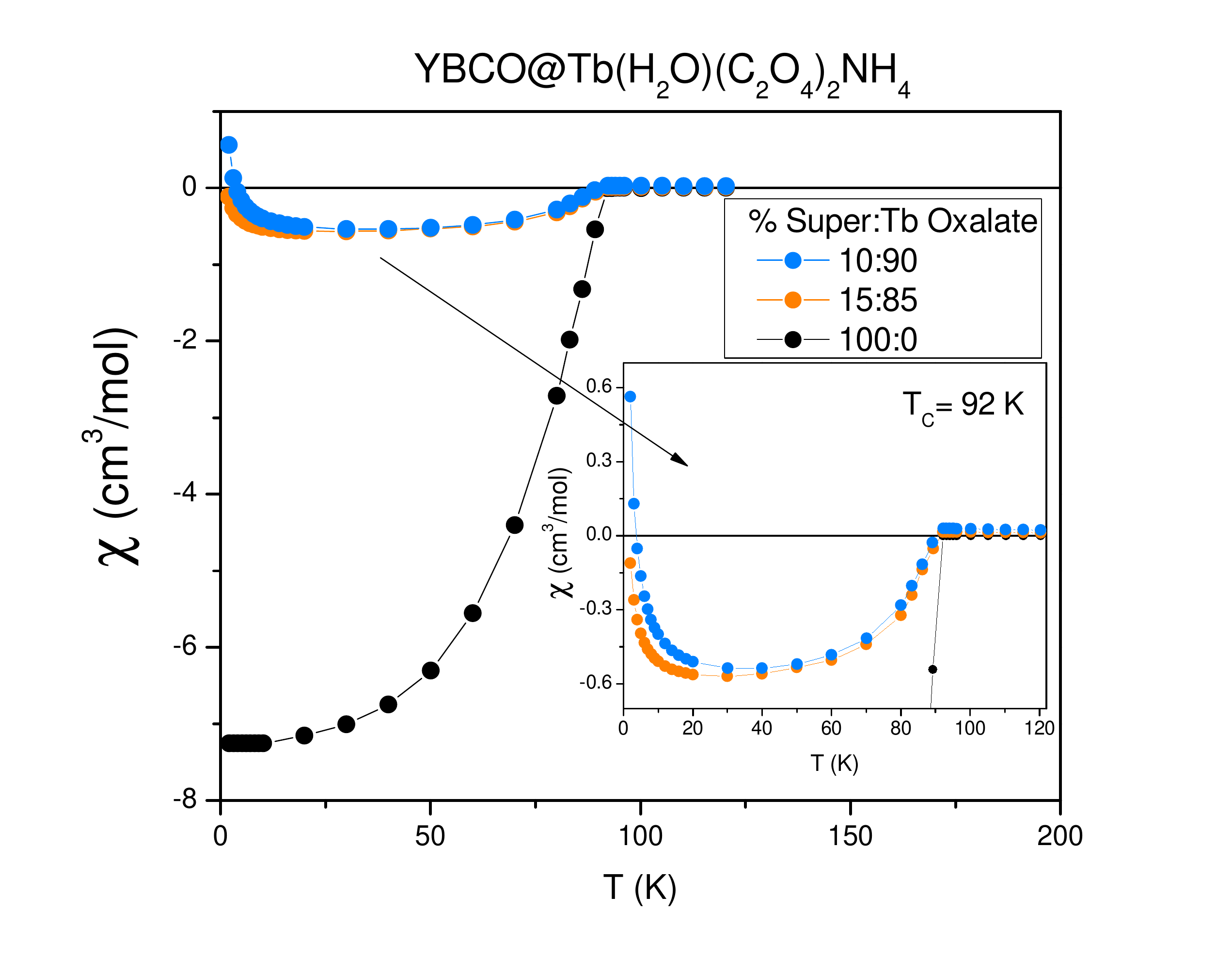}
		   \caption{(Color on-line)  $\chi(T)$ measurements  for the composite  YBCO - Tb(H$_{2}$O)(C$_{2}$O$_{4}$)$_{2}$ $\cdot$ NH$_{4}$ at 100 Oe. The amounts of superconducting material are 10, 15, 100\%.}
	\label{Fig2}
\end{figure}

\subsection{EXPERIMENTAL PROCEDURE}
The materials used  for this investigation were a ceramic superconductor and two different luminescence materials. The  ceramic superconductor, YBCO, presents a high transition temperature, $T_C = 92$ K, and the amount of superconducting material was determined by  two modes of measurements, Zero Field Cooling (ZFC) and Field Cooling (FC) under small magnetic field, 100 Oe.  The  two  luminescence compounds  were Ammonium Terbium Oxalate, Tb(H$_{2}$O)(C$_{2}$O$_{4}$)$_{2}$ $\cdot$ NH$_{4}$, and Yttrium-Europium Oxide, Y$_2$O$_3$:Eu$ ^{3+}$ \cite{Rodo, antic}. The use of these  two compounds is because they are excellent photoluminescence materials.
 
The composites were prepared  using  the  ceramic superconductor and one  luminescent material; YBCO - Tb(H$_{2}$O)(C$_{2}$O$_{4}$)$_{2}$ $\cdot$ NH$_{4}$,  and YBCO -Y$_2$O$_3$:Eu$^{3+}$. Only the composite with  ammonium terbium oxalate gave  a reentrant characteristic.   

  \begin{figure}[h]
	\centering
		\includegraphics[width=0.50\textwidth]{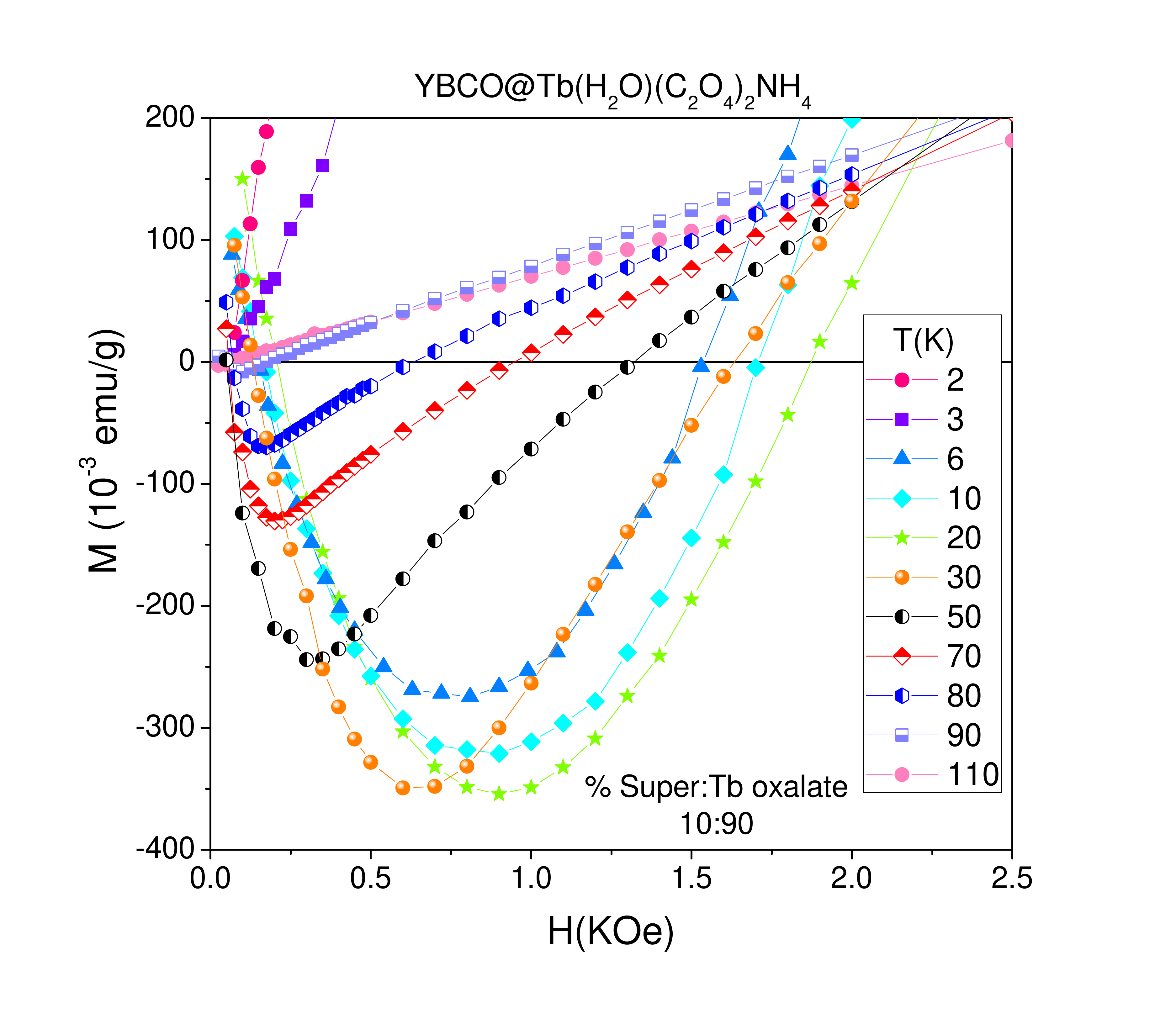}
		   \caption{(Color on-line) Isothermal magnetic measurement, $M-H$,  in mode ZFC from 2 to 110 K for composite  Super:Tb oxalate 10:90\%. This figure shows the decreasing of the critical fields H$ _{C1} $ and H$ _{C2} $ with temperature.  }
	\label{Fig3}
\end{figure}
The preparation of the ceramic superconductor was using a  known procedure  as  already published in the literature, starting from BaCO$_3$, Y$_2$O$_3$, and CuO \cite{Jin}. Subsequently, the  ceramic was grounded and suspended in high pure acetone to evaluate the particle size according to precipitation time. Only we  used particles in  range about  500 and 350 nm.

The composites were prepared by homogeneous precipitation method,  starting from a aqueous solution of rare-earth nitrates, ammonium oxalate and a certain amount of superconducting particles. This  solution was ultrasonically dispersed for 30 minutes in a sealed flask. After, the solution was heated at 75$^{\circ}$ C for two hours with magnetic stirring. The resulting mixture  was washed three times with  30 ml of pure acetone to remove the ammonium oxalate  excess. Finally, the resulting compounds were dried at 60$^{\circ}$ C. Only in the case of YBCO@Y$ _{2} $O$ _{3} $:Eu$ ^{3+} $ composite it was heated at 400$^{\circ}$ C for five minutes.  With this procedure both,  yttrium and europium nitrates,  transforms to oxides.

\begin{figure}[h]
	\centering
		\includegraphics[width=0.45\textwidth]{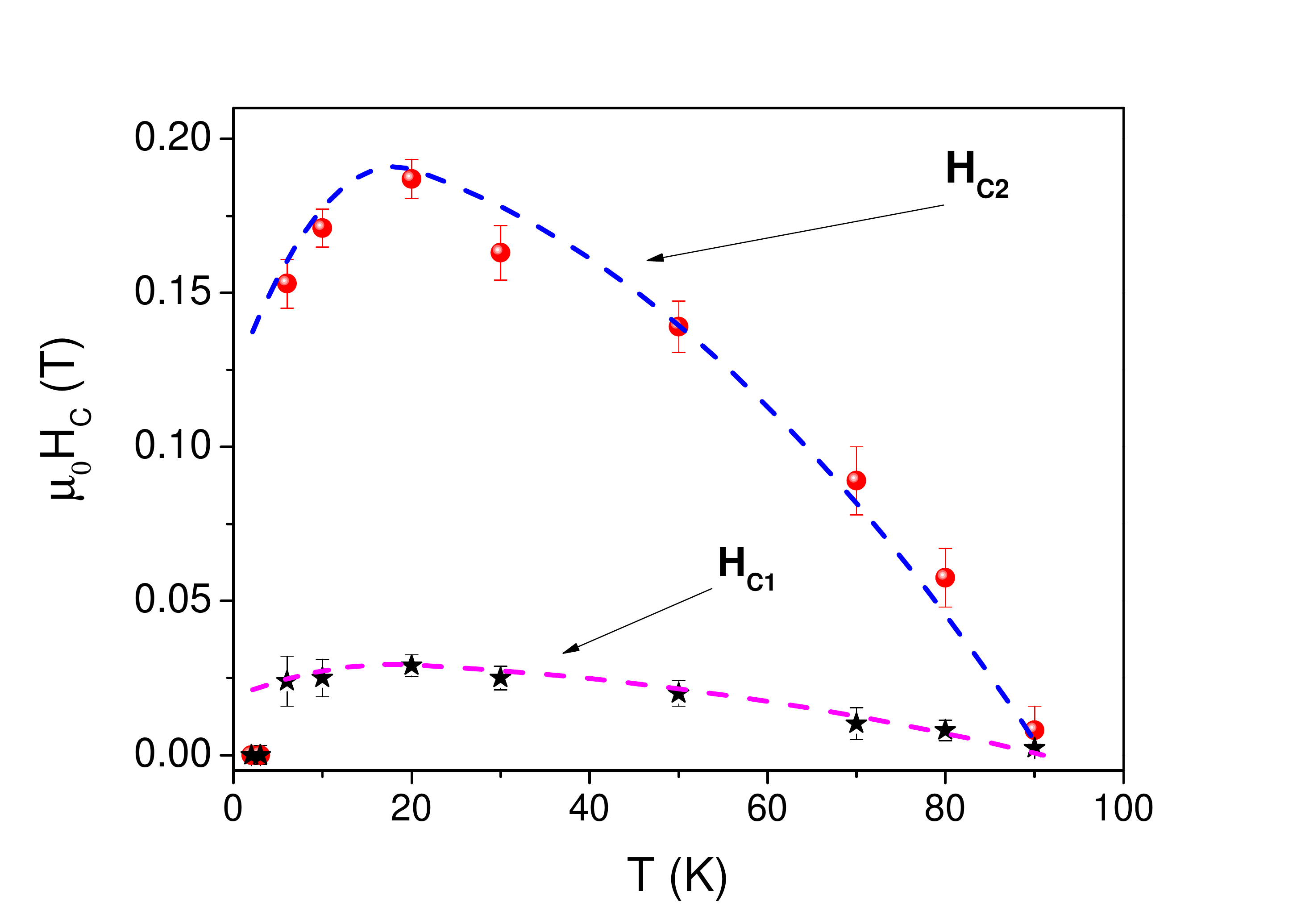}
		   \caption{(Color on-line) The critical fields of one composite with 10\% superconductor and 90\% Tb oxalate. The graphs  were plot using data of Fig 3.  The critical fields were determined in a conventional manner. Note the dramatic reduction and shape of them. This is clearly due to the coexistence between supercondutivity and magnetism in the composite. In this case the calculation of $\kappa$ was quite reduced at a  value around 1.6 in comparison to $\kappa\geq 70$ in YBa$ _{2} $Cu$ _{3} $O$ _{7-\delta} $ single-crystal \cite{Riseman}. The dotted lines are a guide for the eye.}
	\label{Fig4}
\end{figure}

\begin{figure}[h]
	\centering
		\includegraphics[width=0.50\textwidth]{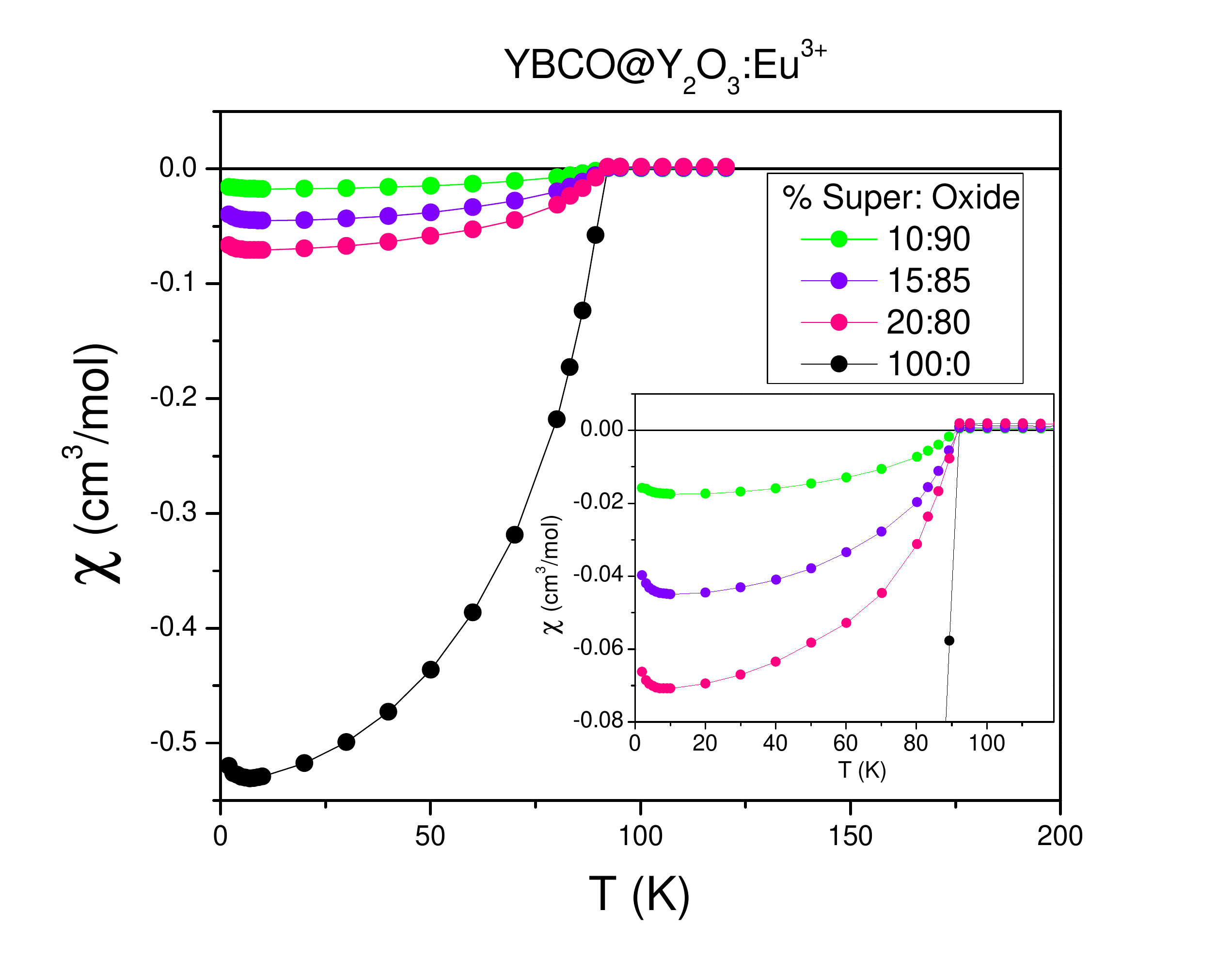}
		   \caption{(Color on-line) Magnetization - Temperature (M−T) in ZFC mode at H=100 Oe in YBCO - Y$_2$O$_3$:Eu$^{3+}$ composite.  This composite is not reentrant as mentioned in the main text. Note also that in this  figure in black dots, the pure ceramic superconductor has  an upturn at low temperatures which is due to  ferromagnetism because surface oxygen vacancies, as was mentioned by other authors \cite{Coey, zhu, li}}. 
	\label{Fig5}
\end{figure}

\begin{figure}[h]
	\centering
		\includegraphics[width=0.40\textwidth]{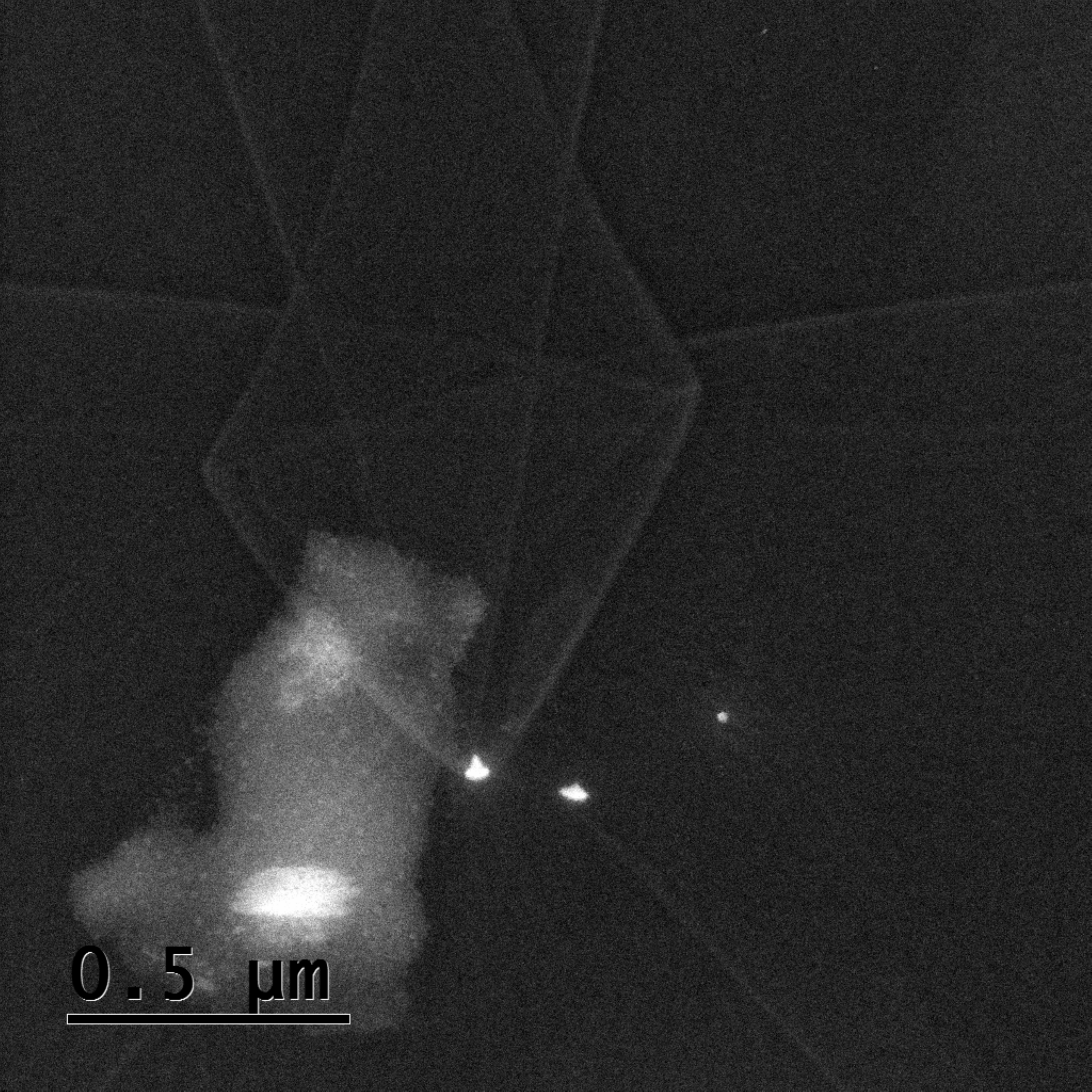}
		   \caption{(Color on-line)  Transmission electron micrography in YBCO - Tb(H$_{2}$O)(C$_{2}$O$_{4}$)$_{2}$ $\cdot$ NH$_{4}$ composite.  White area show the superconducting ceramic, the gray area is the ammonium Tb oxalate.}
	\label{Fig6}
\end{figure}

\begin{figure}[h]
	\centering
		\includegraphics[width=0.5\textwidth]{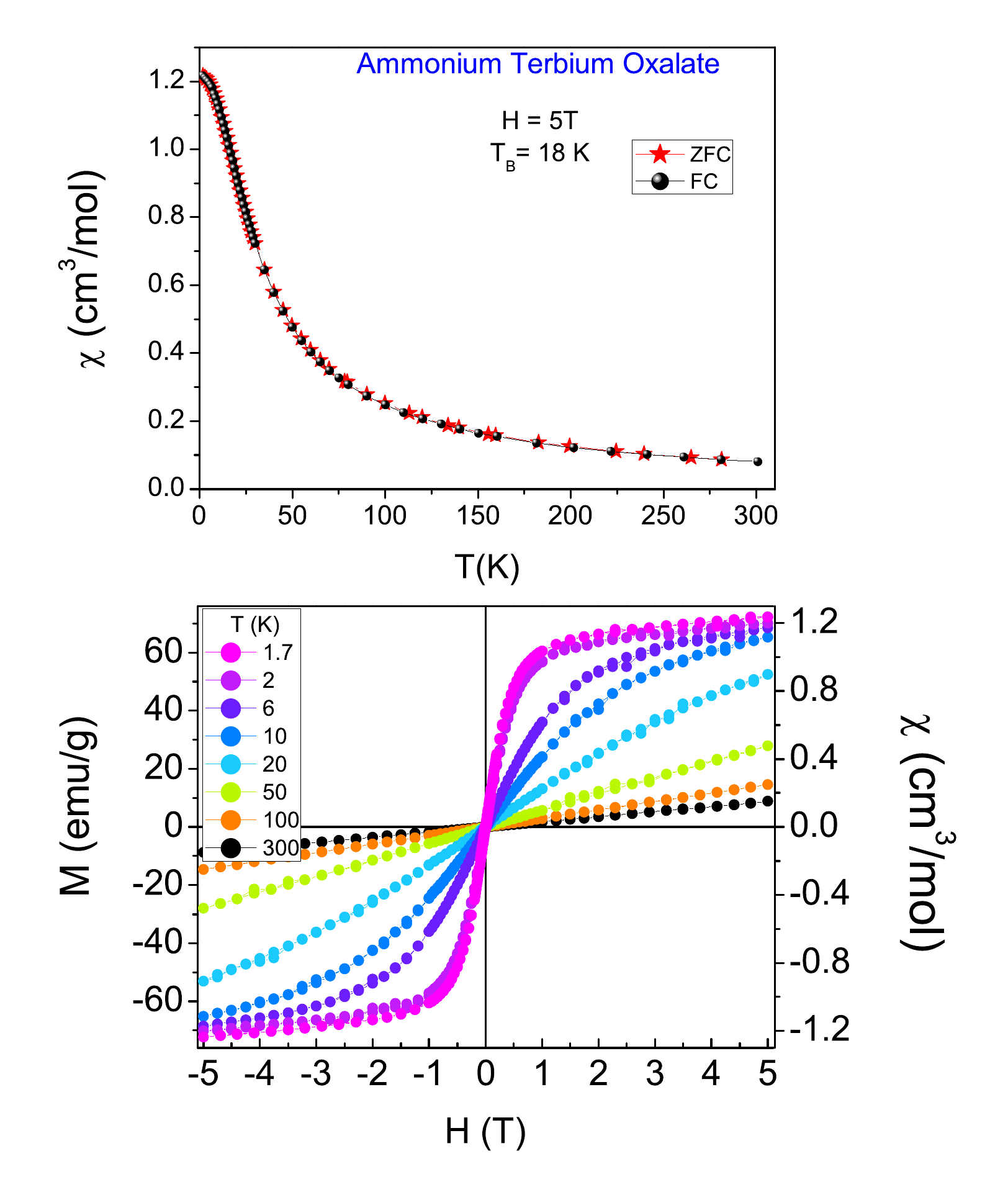}
		   \caption{(Color on-line) Tb(H$_{2}$O)(C$_{2}$O$_{4}$)$_{2}$ $\cdot$NH$_{4}$ magnetic characteristic. Top figure shows $\chi$(T) determines in ZFC and FC with magnetic field of 5 T. The bottom figure shows Magnetization - Magnetic field from  1.7  to 300 K. Note the Tb oxalate shows superparamagnetism with a blocking temperature $ \approx $ 18 K. Right axis shows $\chi$(T), whereas the left axis is in $ M-T$ scale.  }
	\label{Fig7}
\end{figure}

\section{Results and discussion}

 As above mentioned the main idea of this  research was to study the influence of luminescent compounds in the superconducting properties. 

 The characteristics  of the ceramic superconductor are  shown in Fig. 1, top figure. There we also display the characteristics of the composite formed with the Tb oxalate at different compositions Superconductor - Tb oxalate; 10:90 and 15:85 (in weight). In these three figures the data was obtained  in  two modes of measurements, ZFC and FC. At low temperature is seen the reentrant behaviour, both middle and bottom figure show in the two modes of measurements the reentrance characteristic.  In Fig. 2, the main figure and the inset shows only the ZFC measurement. The three curves show three different compositions in the composite. The inset shows that 10, and 15 \% of superconductor are reentrant.       

Figure 3  shows the  isothermal $M-H$ characteristics  from  2 to 110 K for the composite YBCO - Tb(H$_{2}$O)(C$_{2}$O$_{4}$)$_{2}$ $\cdot$ NH$_{4}$  with composition  10:90\%. From this figure  we obtained the variation with temperature and  magnitudes of the critical fields $H_{C2}$ and $H_{C1}$. Fig. 4 presents the behaviour of the critical fields, note the curvature  at low temperatures. These curvatures marks the temperature when the external magnetic field energy  initiate the  breaking  the superconducting behavior.  This shape is the remarkable characteristic in reentrant superconductors  and is when both phenomenons  coexist; superconductivity and magnetism \cite{lyn, bur, Morales}.

It is interesting to mention that these results extracted from Fig. 3 and plotted in Fig. 4 indicate that   the magnitude of the critical fields shows a dramatically reduction,  with magnitudes as small as $ \approx $ 300 Oe and 2 kOe for $H_{C1}$ and $H_{C2}$, respectively. Accordingly, the critical second magnetic field is enormously reduced. This is perhaps the most important discovery in this new reentrant superconducting composite. Evaluating the Ginzburg-Landau parameter from Fig. 4, it was quite reduced at around 1.6 in comparison to the $\kappa\geq 70$ in YBa$ _{2} $Cu$ _{3} $O$ _{7-\delta} $ single-crystal \cite{Riseman}.

On the other hand, in Fig. 5 is shown the $\chi(T)$ characteristic in the YBCO - Y$_2$O$_3$:Eu$^{3+} $ composite. In it  was not seen reentrant behaviour. In this composite we show different proportions of superconducting material from 10 to 100\% and it never becomes reentrant, perhaps because the magnetic response of the Y oxide is smaller than in the Tb oxalate.  An additional observation related to the behavior in the pure ceramic superconductor, black dots of Fig. 5, is that at low temperatures it seen an upturn in $\chi(T)$  which we attributed to  defects and oxygen vacancies on the surface of the superconductor, this has been also observed by other researchers \cite{Coey, zhu, li}.
Lastly  in Fig. 6 is displayed a electron microscopy picture of the composite YBCO - Tb oxalate.  In there, white areas are particles of YBCO, while the grey areas are the Tb oxalate.
 
In order to have a better information of the Tb oxalate, in  Fig. 7 we show the magnetic characteristics. Top figure is the magnetic susceptibility, $\chi - T$ determined at  5 Tesla in  ZFC and FC modes. The bottom figure displays isothermal measurements determined at temperatures from  1.7 to 300 K. Tb oxalate is superparamagnetic with a blocking temperatura $ \approx $ 18 K.

\section{Conclusions}
We found a new reentrant superconductor formed by a Hight-T$ _{c} $ superconductor and a luminescent material. Two composites were prepared with different luminescent compounds, the composite with ammonium terbium oxalate presented reentrant behavior. According to this study  values and shape of the critical magnetic fields, $H_{C1}$ and $H_{C2}$, were  dramatically reduced, implying that the Ginzburg - Landau   parameter  ($\kappa$) has  a small value of 1.6 in the  composite 10:90\%. We concluded   that the reentrant behavior is because the coexistence between superconductivity and magnetism in the composite. As mentioned before the main idea of this study  was to see  the influence of a superconductor in the luminescent properties,  as already published by other authors in \ce{MgB_2} \cite{zhi}, but however the investigation was deviated to study this  new characteristics.

\begin{acknowledgments}

This work was supported by DGAPA-UNAM project IT100217 and CONACyT project 254280. The authors thank to A. Pompa, and A. L\'opez for technical help.   Rodolfo E. L\'{o}pez-Romero thanks to CONACyT-M\'{e}xico for scholarship and Dr. F. Morales for the substantial suggestions.

\end{acknowledgments}

\end{document}